\documentclass[3p,times,procedia]{elsarticle}
\usepackage{nupha_ecrc}
\usepackage{amsmath}
\usepackage[]{hyperref}
\usepackage[]{url}

\volume{00}

\firstpage{1}

\journalname{Nuclear Physics A}

\runauth{}

\jid{nupha}

\jnltitlelogo{Nuclear Physics A}

\usepackage{amssymb}

\usepackage[figuresright]{rotating}

\begin{document}

\begin{frontmatter}



\dochead{XXVIIth International Conference on Ultrarelativistic Nucleus-Nucleus Collisions\\ (Quark Matter 2018)}

\title{Signature of gluon saturation in forward di-hadron correlations at the Large Hadron Collider}


\author[label1]{Giuliano Giacalone}

\author[label2]{Cyrille Marquet}

\address[label1]{Institut de physique th\'eorique, Universit\'e Paris Saclay, CNRS, CEA, F-91191 Gif-sur-Yvette, France}

\address[label2]{CPHT, \'Ecole Polytechnique, CNRS, Universit\'e Paris Saclay, Route de Saclay, F-91128 Palaiseau, France}

\begin{abstract}
We argue that gluon saturation effects may be visible in Large Hadron Collider (LHC) data. 
We look at the centrality dependence of the away-side peak of two-particle correlations measured in p-Pb collisions, using both ALICE data at midrapidity, and LHCb data at forward rapidity.
Once the collective flow-like signal measured at the near-side peak is removed from the correlation function, the centrality dependence of the away-side peak turns out to be much stronger in the forward region than at midrapidity. 
We argue that this is a very specific prediction of the saturation framework, and we compare LHCb data to a state-of-the-art calculation within the Color Glass Condensate theory. 
\end{abstract}


\end{frontmatter}


\section{Introduction: The suppression of the away-side peak in p-A collisions}
The Large Hadron Collider (LHC) has the capability of probing the wavefunctions of protons and nuclei at very small values of $x$, a regime where gluon densities are expected to \textit{saturate} due to nonlinear QCD dynamics.
Such effects are predicted by the Color Glass Condensate (CGC) effective theory, a framework of protons and nuclei overly populated with small-$x$ gluons~\cite{Gelis:2010nm}.
Despite its many phenomenological implications, experimental searches have been, so far, rather unsuccessful in providing a fully conclusive proof of the CGC, and of nonlinear QCD dynamics in general.

A potential smoking gun of gluon saturation is the suppression of the away-side peak ($\Delta\phi=\pi$) of the forward di-hadron correlation function in p-A collisions vs. p-p collisions.
This is a well-known probe that over the years has attracted much attention in the context of d-Au collisions at RHIC~\cite{Albacete:2010pg,Stasto:2011ru,Lappi:2012nh}.
Depletion of angular correlations in p-A collisions is intuitive: 
The natural back-to-back correlation of a hadron pair gets destroyed during the interaction with a target CGC, which occurs through a transverse kick of order of the scale of gluon saturation of the target, $Q_s$, which is larger for a nucleus.
This effect is then enhanced if the hadrons are produced in the region of fragmentation of the projectile (forward region).
Indeed, if a pair of partons of transverse momenta $k_1$ and $k_2$ hadronizes into a pair of hadrons of momenta $p_1=z_1k_1$ and $p_2=z_2k_2$, respectively, the longitudinal momentum fractions of the final-state hadrons read:
\begin{align}
\label{eq:kine}
\nonumber x_1 &= \frac{1}{\sqrt[]{s}}\Bigg(\frac{p_{1t}}{z_1}~e^{y_1}+\frac{p_{2t}}{z_2}~e^{y_2}\Bigg), \\
x_2 &= \frac{1}{\sqrt[]{s}}\Bigg(\frac{p_{1t}}{z_1}~e^{-y_1}+\frac{p_{2t}}{z_2}~e^{-y_2}\Bigg).
\end{align}
At the LHC, then, if particles are correlated at large rapidity, $y\sim4$, one obtains $x_1\sim1$ and $x_2\sim10^{-4}$. 
The projectile is thus \textit{dilute}, i.e., probed at large values of Bjorken $x$, whereas saturation is maximized in the target, which is probed in the \textit{dense} small-$x$ region, and is amenable to a CGC description.

In this work, we argue that the suppression of the away-side peak observed in current LHC p-Pb data provides strong and unique evidence in support of the CGC picture. 

\section{Evidence in experimental data}
\label{sec:2}
Gluon saturation emerges in a low-$x$ environment, therefore, any phenomenological signature of the CGC should smoothly disappear as $x$ becomes large [or if rapidity, $y=\ln(1/x)$, goes towards zero].
In this section, we show that a rather intriguing dependence with rapidity, which supports the CGC scenario, is present in the LHC p-Pb data.
The quantity we shall be dealing with is the following normalized yield:
\begin{equation}
Y(\Delta\phi) = \frac{1}{\Delta y}\frac{N^{\rm pair}(\Delta\phi)}{N^{\rm trig}},
\end{equation}
where $N_{trig}$ denotes the number of trigger particles and $N_{pair}$ counts the total number of pairing with an associated particle separated by a relative azimuthal angle $\Delta\phi$. $\Delta y$ is the width of the rapidity interval over which particles are produced.
The generic prediction of the CGC framework is, then, that $Y(\pi)$ is larger in p-p collisions than in p-Pb collisions, and that the discrepancy should be enhanced if particles are correlated at large $y$.
We deal with measurements of $Y(\Delta\phi)$ performed by the ALICE collaboration~\cite{Abelev:2012ola} at midrapidity, and by the LHCb collaboration~\cite{Aaij:2015qcq} in the forward region.
Since data are available only for p-Pb collisions, instead of pointing out a difference between p-p and p-Pb, we shall infer signals of gluon saturation through comparisons between peripheral p-Pb collisions (which are good proxies for p-p collision), and central p-Pb collisions.

The ALICE collaboration reported measurements of $Y(\Delta\phi)$ in p-Pb collisions at midrapidity in different centrality classes~\cite{Abelev:2012ola}.
They estimate the correlation induced solely by effects \textit{unrelated} to collective flow-like phenomena. 
To this aim, they carry out their measurements implementing a large gap in rapidity between correlated particles, which allows to spot, in particular on the near side, flow-like signals of long-range nature.
The resulting shape of $Y(\Delta\phi)$ is then fitted with a $\cos( 2\Delta\phi)$ modulation,
\begin{figure}[t!]
\centering
\includegraphics[width=\linewidth]{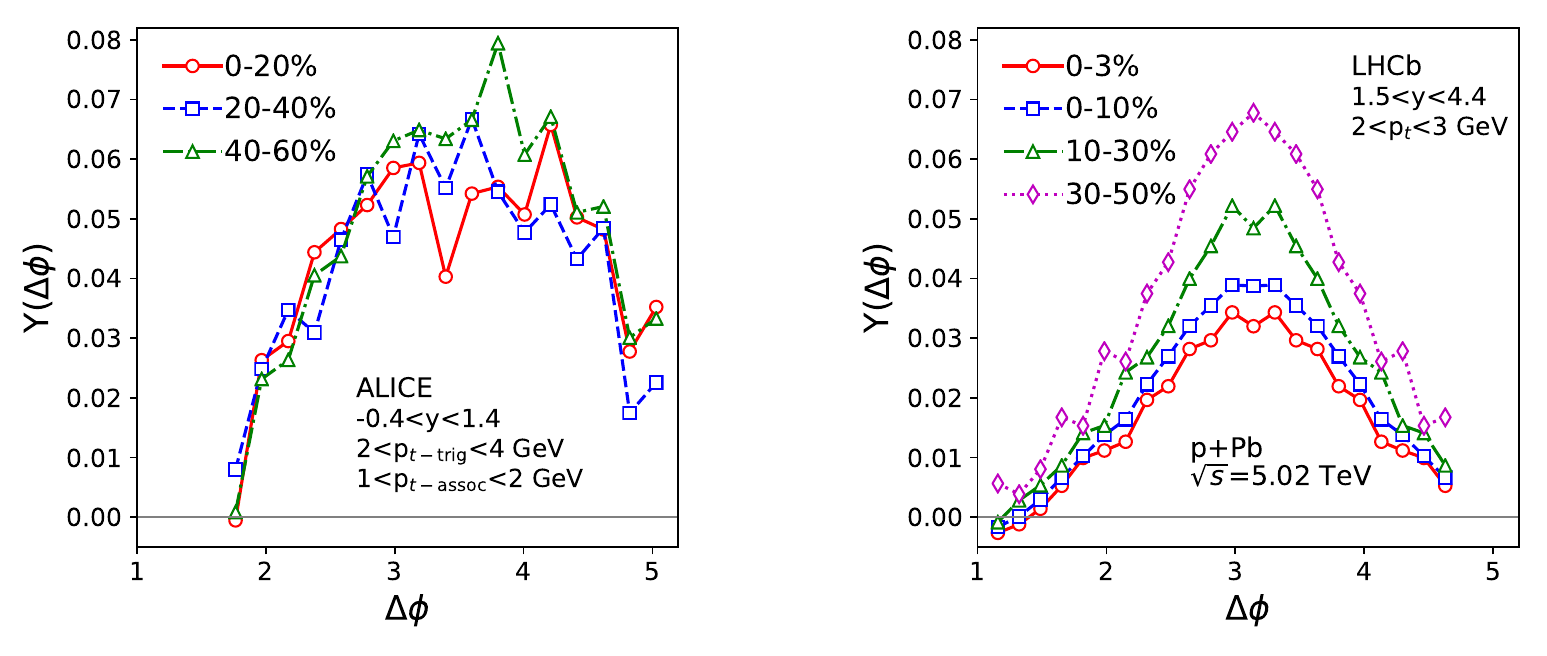}
\caption{Left panel: Centrality dependence of the flow-subtracted per-trigger yield at midrapidity, as function of $\Delta\phi$, measured by the ALICE collaboration in p+Pb collisions at ~$\sqrt[]{s}=5.02$~TeV. Data points extracted from Fig.~5 of Ref.~\cite{Abelev:2012ola}. Right panel: Same but for LHCb data measured at forward rapidity~\cite{Aaij:2015qcq}. In both plots, absolute errors on the points (not shown) are approximately equal to 0.01.}
\label{fig:1}
\end{figure}
\begin{equation}
\label{eq:fit}
Y(\Delta\phi) = a_0 + 2~ a_2 \cos (2 \Delta\phi).
\end{equation}
The fitted "double-ridge" curve is eventually subtracted from the measurement of $Y(\Delta\phi)$, and we show the remaining flow-subtracted signal in the left panel of Fig.~\ref{fig:1}, on which we shall comment in a moment (we use that terminology regardless of what is actually at the origin of the double-ridge structure, which may not be collective flow).

Now, let us perform the same trick on LHCb data~\cite{Aaij:2015qcq}.
We take $Y(\Delta\phi)$, measured with a rapidity gap, in different centrality classes, and we fit the near-side region using Eq.~(\ref{eq:fit}) in order to estimate, at each centrality, the correlation due to long-range phenomena.
An example of such fit, for the 0-3\% centrality range, is given by the dashed line in Fig.~\ref{fig:2}.
Subtracting the near-side contribution from each peak on the away side, we obtain the curves shown in the right panel of Fig.~\ref{fig:1}.

The difference between flow-subtracted ALICE and LHCb data is striking: An evident hierarchy with centrality emerges in forward LHCb data, which is much less visible at midarpidity in the ALICE measurement.
We conclude that the two crucial features predicted by the CGC formalism are both visible in the experimental data: The away-side peak is suppressed in central collisions, and this suppression fades away as we correlate particles away from the forward region.
This is our main result, and is, arguably, the clearest signature of gluon saturation at play in experimental data so far observed at the LHC.

\section{A state-of-the-art CGC calculation}
\label{sec:3}
It is rather legitimate to ask whether the previous data, which support the qualitative CGC predictions, can be reproduced by an actual calculation.
In this section, we perform a calculation of $Y(\Delta\phi)$ for di-hadrons produced back-to-back.
This configuration allows us to exploit the state-of-the-art machinery of di-hadron production, in the so-called transverse momentum dependent (TMD) limit of the CGC.
The TMD cross section for back-to-back production in a generic dilute-dense collision can be packed into the following formula~\cite{Dominguez:2011wm}:
\begin{equation}
\label{eq:tmdf}
\frac{ d \sigma^{pA\rightarrow hh X}}{ d^2 k_{1t}  d^2 k_{2t}  d y_1  d y_2} = \frac{\alpha_s^2}{(x_1 x_2 s)^2} \sum_{a,c,d} x_1 f_{a/p} (x_1,\mu^2) \sum_{i} \frac{1}{1+\delta_{cd}} H_{ag\rightarrow cd}^{(i)}(z,P_t)\mathcal{F}_{ag}^{(i)}(x_2,k_t),
\end{equation}
where
\begin{equation}
k_t = k_{1t} + k_{2t}, \hspace{40pt}z=\frac{k_1^+}{k_1^+ + k_2^+},\hspace{40pt} P_t=(1-z)k_{1t}-zk_{2t},
\end{equation}
and $a$, $c$, $d$ sum over production channels ($qg$, $q \bar q$, $gg$).
We note that the cross section is nicely factorized into a a dilute component, described by parton distribution functions, $f_{a/p}(x_1,\mu^2)$, computed at factorization scale $\mu^2$, and a channel-dependent dense counterpart which is a combination of TMD gluon distributions, $\mathcal{F}^{(i)}(x_2,k_t)$, weighted by the corresponding hard factors~\cite{Dominguez:2011wm}, $H(z,P_t)$.
The TMD gluon distributions give, roughly speaking, the gluon content of the target CGC, and one has to resort to small-$x$ QCD evolution equations (e.g. JIMWLK~\cite{Marquet:2016cgx}, or rcBK evolution~\cite{vanHameren:2016ftb,Albacete:2018ruq}) to obtain them at any small value of $x_2$.
We refer to Ref.~\cite{Albacete:2018ruq} for an exhaustive description of all the ingredients appearing in the cross section.
\begin{figure}[t!]
\centering
\includegraphics[width=.65\linewidth]{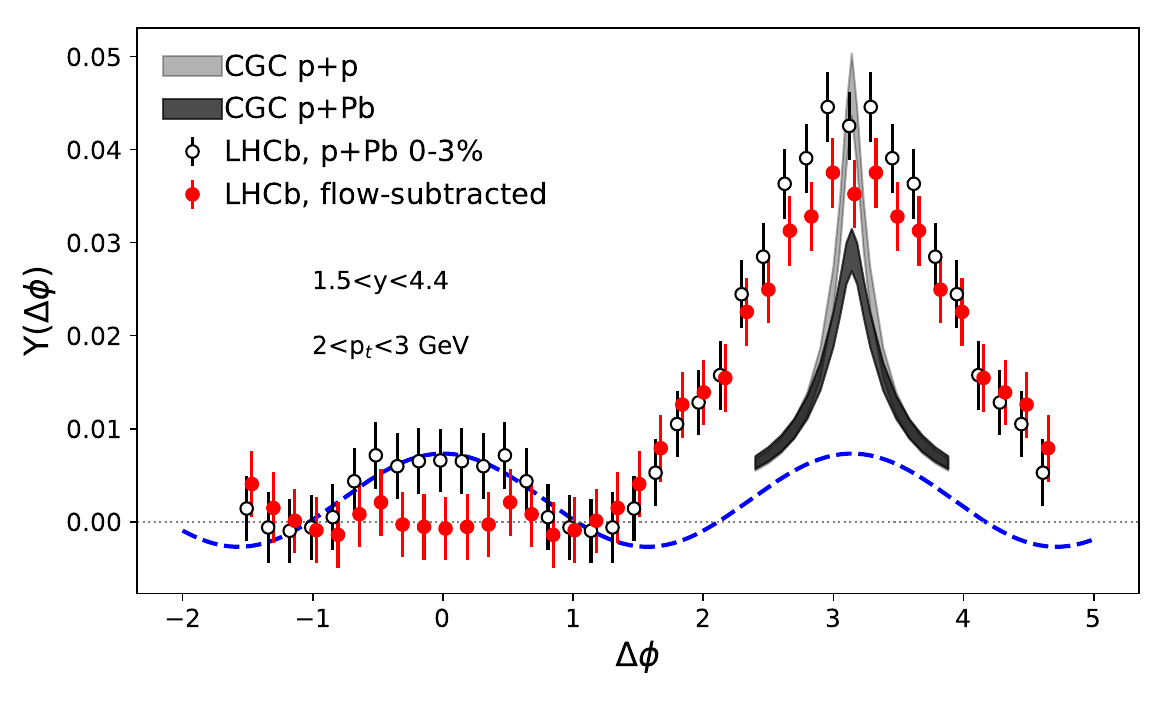}
\caption{Empty Symbols: LHCb data on $Y(\Delta\phi)$ in $p$+Pb collisions at $~\sqrt[]{s}=5~{\rm TeV}$, in the 0--3\% centrality range. Red symbols: LHCb data after subtraction of the fitted near-side peak. The dashed line displays the fit of Eq.~(\ref{eq:fit}) to the flow-like signal. The shaded bands represent CGC results for p-p collisions (light band), and p-Pb collisions (dark band). Both data and theory implement a rapidity gap of 2 units of $y$. Shaded bands indicate the effect of a 50\% variation in the factorization scale, $\mu^2$.}
\label{fig:2}
\end{figure}

Convolving Eq.~(\ref{eq:tmdf}) with fragmentation functions, integrating over the kinematic range of LHCb data, and normalizing by the inclusive single hadron yield~\cite{Albacete:2010bs}, we obtain the results shown as shaded bands in Fig.~\ref{fig:2}.
We see that, when particles are back-to-back, our results for p-Pb match to a good extent the measured $Y(\pi)$, and we predict a visible suppression of the away-side peak as we move from p-p to p-Pb.
Note that this suppression is smaller than typically predicted by CGC calculations~\cite{Albacete:2018ruq}.
The reason is that, as done in the experimental analysis, we implement a gap of two units of rapidity between the correlated hadrons, so that one hadron has much smaller rapidity than the other, and this somewhat reduces the dilute-dense asymmetry, minimizing the effects due to saturation in the target.

Two further comments are in order.
First, these results are obtained by mere application of the formulas used in Ref.~\cite{Albacete:2018ruq}, so that we are considering only fragmentation of quarks and gluons to neutral pions.
Our results include, then, a (rather large) factor $3$ to account for all charged particles, as measured by LHCb, thank to which we are able to match within errors the flow-subtracted data (red symbols). 
This implies that if we had not carefully removed the correlation observed at the near side, we would have ended up underestimating the measured $Y(\pi)$ (white symbols) by a large amount.
Second, as already noted in Ref.~\cite{Albacete:2018ruq}, our peak around $\Delta\phi=\pi$ is much narrower than the data.
The reason is that our calculation is still incomplete, as we do not include radiation of initial- and final-state soft gluons, which would naturally yield a broader correlation function~\cite{Stasto:2018rci}.
This feature will be added to the formalism in a future work.

\section{Summary}
The data-driven analysis carried out in this paper provides convincing evidence of the presence of gluon saturation effects in current LHC data.
We hope this study will motivate further analysis of the rapidity dependence of the suppression of the away-side peak, in particular, it will be crucial to perform the same measurements in p-p collisions, possibly at the same center-of-mass energy.
The state-of-the-art TMD calculation leading to the results of Fig.~\ref{fig:2} proves able to describe the measured correlation at $\Delta\phi=\pi$, provided that one subtracts from the data the flow-like signal observed on the near side.





\bibliographystyle{elsarticle-num}
\bibliography{biblio}







\end{document}